\documentclass{emulateapj}

\newcommand{\sax}{SAX~J1808.4$-$3658}
\newcommand{\saxs}{J1808.4}

\begin{document}

\title{An Accurate Determination of the Optical Periodic Modulation 
in the X-Ray Binary SAX J1808.4$-$3658} 

\author{Zhongxiang Wang\altaffilmark{1}, 
Cees Bassa\altaffilmark{2,3}, 
Andrew Cumming\altaffilmark{1}, 
and Victoria M. Kaspi\altaffilmark{1}}

\altaffiltext{1}{\footnotesize Department of Physics,
Ernest Rutherford Physics Building,
McGill University, 3600 University Street, Montreal, QC H3A 2T8, Canada}

\altaffiltext{2}{\footnotesize SRON Netherlands Institute for Space Research,
Sorbonnelaan 2, 3584 CA, Utrecht,
The Netherlands}

\altaffiltext{3}{\footnotesize Department of Astrophysics,
IMAPP, Radboud University Nijmegen, 
Toernooiveld 1, 6525 ED, Nijmegen,
the Netherlands}

\begin{abstract}
We report on optical imaging of the X-ray binary \sax\ with the 8-m Gemini 
South Telescope. The binary, containing an accretion-powered millisecond 
pulsar, appears to have a large periodic modulation in its quiescent optical 
emission.  In order to clarify the origin of this modulation, 
we obtained three time-resolved $r'$-band light curves (LCs) of the source
in five days.  The LCs can be described by a sinusoid, and 
the long time-span between them allows us to determine 
optical period $P=7251.9$~s and phase 0.671 at
MJD 54599.0 (TDB; phase 0.0 corresponds to the ascending node 
of the pulsar orbit), with uncertainties of 2.8~s and 0.008 
(90\% confidence), respectively.
This periodicity is highly consistent with the X-ray orbital ephemeris.
By considering this consistency and the sinusoidal shape of the LCs, we rule 
out the possibility of the modulation arising from the accretion disk.
Our study supports the previous suggestion that the X-ray pulsar becomes
rotationally powered in quiescence, with its energy output irradiating
the companion star, causing the optical modulation. 
While it has also been suggested that the accretion disk would be
evaporated by the pulsar, we argue that the disk exists
and gives rise to the persistent optical emission.
The existence of the disk can be verified by long-term, multi-wavelength 
optical monitoring of the source in quiescence, as an increasing
flux and spectral changes from the source would be expected based 
on the standard disk instability model.

\end{abstract}


\section{INTRODUCTION}

While it had long been believed that neutron star (NS) X-ray binaries (XRBs) 
are progenitors of the recycled millisecond radio pulsars \citep{bv91},
it was the discovery of coherent pulsations from the transient XRB 
\sax\ (hereafter \saxs) during its X-ray outburst in 1998 that first 
and finally confirmed the connection between the two systems \citep{wv98}:
in this binary, the accreting NS is a 2.49 ms X-ray pulsar. 
As the first example of accretion-powered millisecond pulsar systems, 
\saxs\ has been extensively studied, with various interesting properties 
revealed (see \citealt{har+08} and references therein).  In this paper, 
we focus on the optical periodic modulation seen in this binary and
report on our observational study of the modulation.

The orbital period of \saxs\ is $P_{\rm orb}\simeq$7249.157 s 
($\simeq$2.01 hr), accurately known to one part in $10^{10}$ 
from Doppler modulations of 
the millisecond pulsations \citep{cm98,har+08}.
Combined with the derived mass function of 3.8$\times 10^{-5} M_{\sun}$, 
the period implies that the mass-transferring companion could be a 
0.17 $M_{\sun}$ low-mass main-sequence star, but more likely
a $\sim$0.05 $M_{\sun}$ brown dwarf \citep{bc01}.
At a distance of $D=3.5$ kpc \citep{gc06}, the optical counterpart
in quiescence is several magnitudes brighter ($V=20.7$,
$L_{V}\simeq 3.0\times 10^{32}$ ergs s$^{-1}$ assuming isotropic emission
and extrinction $A_V=0.73$; see \S~2 and \S~4) 
than the possible types of stars suggested as
the companion, probably indicating that the optical emission
arises from the accretion disk in the binary \citep{hccv01}. 
However in the quiescent state,
10--40\% sinusoidal-like modulations in the source's 
optical light curves (LCs) have been reported (\citealt{hccv01,cam+04}), 
and this is puzzling because the quiescent X-ray 
luminosity is approximately $L_{\rm X}\simeq 5\times 10^{31}$ ergs s$^{-1}$
(e.g., \citealt{hei+07}), two orders of magnitude lower than that 
required to account for the modulation \citep{bur+03}. 
Typically in a low-mass X-ray binary (LMXB), 
sinusoidal optical modulation arises from X-ray heating of 
the companion star by the central X-ray source: the visible area 
of the heated face varies as a function of 
orbital phase (e.g., \citealt{ak93}).  In \saxs, depending on 
the companion's star types, only 0.5--1.4\% [estimated by
$(R_2/D_b)^2/4$, where $R_2$ is radius of the companion and
$D_b$ is the separation distance between the NS and companion] of 
the total energy flux from the central NS would be received by 
the companion for isotropic emission.
This leads to the suggestion that in quiescence, the NS might switch to be 
a rotation-powered pulsar so that the rotational energy
would be the energy source that heats half surface of the companion star 
and causes the modulation \citep{bur+03}.

However, there are other possibilities that do not require
a rotation-powered pulsar, and we have considered whether or not
the accretion disk could give rise to the modulation.
It has been known that ``superhumps", which are commonly seen 
in short-period cataclysmic variables (CV; \citealt{war95}), 
also appear in LMXBs (e.g., \citealt{has+01}).
These periodic modulations have periods a few percent longer
than the orbital periods and can be sinusoidal-like with an amplitude
of $\sim$10\%, arising from a precessing, 
eccentric accretion disk (e.g., \citealt{wk91}).
Indeed, it has been suggested that those NS LMXBs 
with $P_{\rm orb} <4.2$ hr are potential superhump sources \citep{has+01}. 
In addition, several parts of an accretion disk could  
contribute significantly to optical modulation (e.g., \citealt{mc82}). It
has also been suggested that for an X-ray transient, its quiescent optical 
emission may come from a bright spot on the accretion disk \citep{mc01}. 

\begin{figure}
\plotone{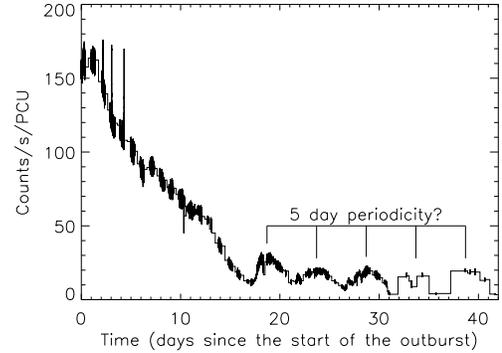}
\figcaption{X-ray light curve of \saxs\ during its 2002 outburst, 
obtained with the Proportional Counter Array (PCA; 2--60 keV energy range) 
on board the \textit{Rossi X-ray Timing Explorer} satellite. Gaps in 
the light curve were due to Earth occultations of the source. At the end of 
the outburst, a $\sim$5 day periodicity is tentatively suggested.
\label{fig:xlc}}
\end{figure}

In particular, the superhump possibility was suggested by 
the X-ray LC obtained in the source's 2002 outburst.
As shown in Figure~\ref{fig:xlc}, the LC exhibits a $\sim$5-day periodic 
modulation at the end of the outburst. If this indicates the precession 
periodicity ($P_{\rm prec}\simeq 5$ days) of the accretion disk, 
it would imply a superhump period of $P_{\rm sh}= 7373$~s 
($1/P_{\rm sh}=1/P_{\rm orb}-1/P_{\rm prec}$) and superhump excess 
$\epsilon =0.017$ 
[$\epsilon =(P_{\rm sh}-P_{\rm orb})/P_{\rm orb}$] \citep{pat01}.  
The excess value is consistent with those obtained for cataclysmic variables
and LMXBs (\citealt{pat+05, has+01}). Furthermore, 
a mass ratio of $q\simeq 0.08$ could be estimated from the
relation $\epsilon =0.18q+0.29q^2$ \citep{pat+05}, implying a companion mass of
0.11 $M_{\sun}$ for 1.4 $M_{\sun}$ NS mass. This companion mass is  
within the range implied by the mass function.

Previously, time-resolved imaging observations over a small period 
of time (covering only $\sim$1.5 orbital period of the binary) were made.
However, these observations were carried out either with a small 
telescope (\citealt{hccv01}) or under very poor observing conditions
\citep{cam+04}, resulting in large uncertainties in the obtained LCs. 
In order to study the optical emission from \saxs, and particularly to
probe whether it could be a superhump source, we have 
obtained high quality optical LCs of the source 
in its quiescent state through time-resolved photometry. 
The observations were made with the 8-m Gemini South Telescope 
over five days, allowing us to determine the
period and phase of the optical modulation accurately. 
We note that \citet{hei+08} (see also \citealt{del+08}) recently 
observed the source simultaneously 
at X-rays and optical $g'i'$ wavelengths, and from the observations 
they confirmed the inconsistency between the large amplitude optical modulation 
and low X-ray luminosity. 

\section{OBSERVATIONS AND DATA REDUCTION}    
\label{sec:obs}

To determine the periodicity in the source's optical emission accurately, 
three Gemini queue mode observations of \saxs\ were carried out in five days, 
on 2008 May 11, 12, and 15. The starting time of each observation
was approximately 06 hour (UTC) each day, resulting in a time span of 
$\simeq$4 days between the first and third observations.
We proposed such observations because we estimated that the time span 
would allow us to determine the period to $<$10 s accuracy, and the second 
observation would be needed to keep the track of the optical periodicity phase.
A Sloan $r'$ filter, with the central wavelength at 6300 \AA, was used
for imaging.  The detector was the Gemini Multi-Object 
Spectrograph (GMOS; \citealt{hoo+04}), which consists of 
three 2048$\times$4608 EEV CCDs.  We used $2\times2$ binning, 
providing a pixel scale is 0.146\arcsec\ pixel$^{-1}$. 
In each night,
36 images of the source were obtained contiguously, each with 
approximately 3.9 min exposure time. The detector's slow read mode, 
having 55~s readout time, was used.  As a result, the total observation 
time in each night was approximately 3 hrs, covering 1.5 orbital 
cycles of \saxs. The average seeing [full-width half-maximum (FWHM) of 
the point spread function (PSF) of the images]
for the three nights were 0.63\arcsec, 0.58\arcsec, and 0.70\arcsec, 
respectively. The second night had the best observing conditions, 
with the seeing reaching 0.51\arcsec\ a few times during the observation.

We used the Gemini IRAF package {\tt GMOS} for data reduction. The images
were bias subtracted and flat fielded. The bias and flat frames
were from GMOS baseline calibrations, made on 2008 May 13 and
May 11, respectively. The standard star used for flux calibration
was PG1047+003A \citep{smi+02}. The observation of this star was
made on 2008 May 13, also as part of the GMOS baseline calibrations.
The airmass of the observation was 1.234, which can be estimated to 
have caused a zero-magnitude offset of 0.03 mag\footnote{www.gemini.edu/sciops/instruments/gmos/calibration/photometric-stds}. 
We did not add this offset to our brightness measurements given below; instead
we consider it as an uncertainty for absolute flux calibration.

\begin{figure}
\plotone{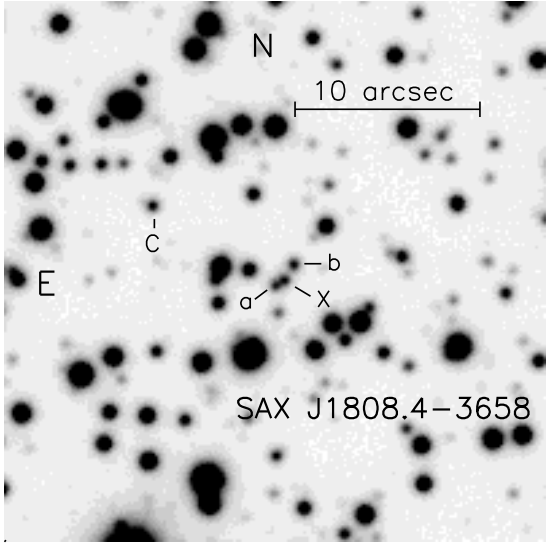}
\figcaption{Gemini South $r'$ image of the \saxs\ field.
Object $X$, located between star $a$ and $b$, is the optical counterpart
to \saxs. The star labeled as $C$ is used as a check star. 
\label{fig:finding}
}
\end{figure}

We performed PSF-fitting photometry to measure the brightnesses of 
the source and other in-field stars.  A photometry program 
{\tt DOPHOT} \citep{sms93} was used.  A finding chart of the target is 
shown in Figure~\ref{fig:finding}.
As can be seen, our target is located between two stars with similar 
brightnesses. Its distance to star $a$ is 0.6\arcsec\ and to star $b$
is 1.0\arcsec. In a few of images, we have FWHM around 0.8\arcsec;
in these cases, our target and star $a$ are nearly unseparated. For these
images, we positionally calibrated them to a reference image that 
was combined from four best-quality images in night 2.
We determined the positions of our target and star $a$ in the reference image 
and fixed them at the positions for photometry of the images. 

We performed differential photometry to eliminate systematic flux variations 
in the images. An ensemble of 8 isolated, nonvariable stars in the 
field were used. The brightnesses of our target and other stars in each
image were calculated relative to the total counts of these stars. Star $C$ 
(Figure~\ref{fig:finding}) was used as a check star, because it was 
non-variable and had similar brightness to our target.

We used the third image from the second night to obtain absolute magnitudes of 
the target and nearby stars, as it is one of the best-quality images.
The aperture correction was calculated
using 15 in-field stars, with an uncertainty of 0.025 mag.
The resulting magnitudes of the target are given in Table~\ref{tab:sax},
and the average magnitudes of the nearby stars $a$ and $b$, and
the check star $C$ were $r'=21.492\pm0.048$, 21.133$\pm$0.013,
and 21.178$\pm0.013$ mag, respectively. The LCs of our target and 
stars $a$ and $C$
are shown in Figure~\ref{fig:lc}. As can be seen, star $a$ was likely
a variable, with the magnitudes and standard deviations of its three
LCs being $21.545\pm0.029$, $21.444\pm0.020$, and $21.486\pm0.021$.
The difference between the first and second nights is
2.9$\sigma$ significant. These results are summarized in 
Table~\ref{tab:pho}.

As we compared our results with those previously reported,
we noted that the source magnitudes, resulting from imaging 
observations made on 1999 July 11 with the Very Large Telescope (VLT) at 
the European Southern Observatory, 
are approximately 1 mag lower than the values
given by \citet{cam+04}, who analyzed the same data. 
The data consist of 1\,min exposures in the $V$, $R$, and $I$-bands, taken with 
the high resolution collimator, providing a pixel scale of 
0.1\arcsec\ pixel$^{-1}$.
The instrumental magnitudes were calibrated against photometric
standard stars in the SA110 field \citep{lan92}. We
obtained $V=20.73\pm0.04$, $R=20.59\pm0.04$, and $I=20.15\pm0.06$, 
where the uncertainty is the quadratic sum of the uncertainty in the
zeropoint, the aperture correction, and the instrumental magnitude.
Comparing the magnitudes of the in-field stars,
including star $a$ and $b$, from the VLT observations
and ours, we believe that our magnitude values are correct.

\section{Periodicity Determination}

\begin{figure*}
\plotone{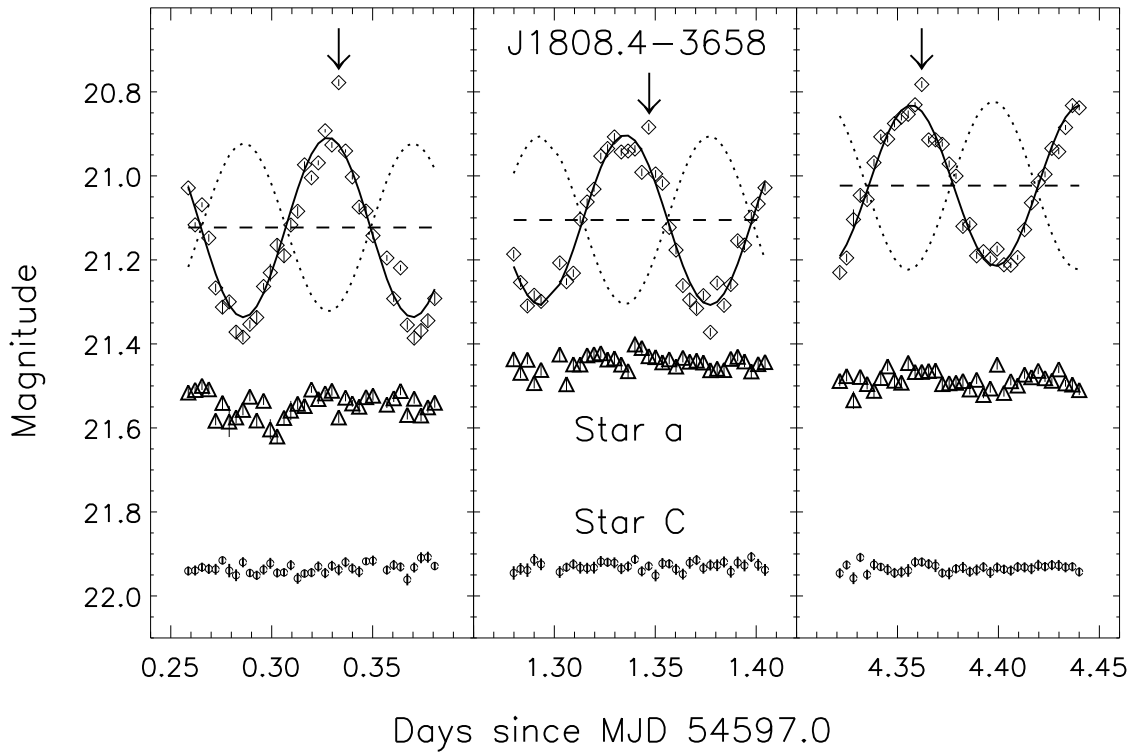}
\figcaption{Optical $r'$ light curves of \saxs\ (diamonds), in which sinusoidal 
modulation is clearly visible. For comparison, the LCs of the nearby star $a$
(triangles) and check star $C$ (circles, downward shifted by 0.8 mag)
are also shown. The best-fit sinusoidal function to the LCs of \saxs\ 
is shown as the solid curves, while the constant magnitude for each LC is 
indicated by the dashed lines. The optical periodicity well matches the 
X-ray ephemeris (dotted curves), which gives the mean 
orbital longitude of the binary \citep{har+08}. The optical brightness peaks 
correspond to when the pulsar is right in front of the companion 
star (270\arcdeg\ mean orbital longitude). The brightest data points
in the LCs, indicated by arrows, are at phase 0.05-0.17 after the maxima
of the sinusoid.
\label{fig:lc} }
\end{figure*}

As can be seen in Figure \ref{fig:lc}, the LCs of \saxs\ clearly show 
a sinusoidal modulation, and appear to have different average brightnesses,
indicating overall variations from day to day. The times of the data points
are barycentric corrected, with the JPL Solar System Ephemeris
DE405 used. In order to determine the modulation, we fit the LCs
with function $m=m_c + m_h\sin [2\pi (t/P + \phi_0)]$, where $t$ is 
the time,  $P$, $\phi_0$, and $m_h$ are the period, starting phase, 
and semiamplitude of the sinusoidal modulation, respectively. 
The parameters $m_c$ and $m_h$ were kept as a constant for each LC, but were 
allowed to have different values in different LCs.
As a result, we found that the best-fit sinusoid 
($\chi^2=1879$ for 100 degrees of
freedom) has $P=7251.9$ s and $\phi_0=0.671$ at MJD 54599.0 (TDB;
Phase $\phi =0.0$ corresponds to the ascending node of the pulsar orbit).

While the LCs can be described by the sinusoidal function,
as shown in Figure~\ref{fig:lc},
the large $\chi^2$ value indicates large scattering of the data
points from the best-fit function. There is a systematic uncertainty
caused by our target's proximity to star $a$. This can be seen from 
the fact that the standard deviations of the three LCs of star $a$ are
significantly larger than its uncertainties from
PSF-fitting (the average is 0.013 mag) and the standard
deviation (0.013 mag) of all data points of the check star $C$.
In addition, we also independently used the program {\tt DAOPHOT} in the 
ESO-MIDAS system for photometry. The resulting LCs are
very similar to those resulting from {\tt DOPHOT},
but with the standard deviations of the differences between
the two sets of the LCs being 0.027, 0.019, and 0.014 mag for the three nights.
These values are approximately consistent with the standard deviation values
of star $a$, confirming the contamination of the photometry 
caused by the proximity
of our target and star $a$. Adding the standard deviations of star $a$ in
quadrature with the uncertainties of data points (resulting from PSF-fitting)
of the target, the $\chi^2$ value is reduced to 266 for 100 degrees of
freedom. This indicates that there is intrinsic scattering of 
the data points from the single sinusoid. For example, we note that 
the brightest data point
in each LC appears at phase 0.05--0.17 after the maximum of the sinusoid.
This pattern is likely to be true, because the {\tt DOPHOT} 
and {\tt DAOPHOT} measurements at the LCs' region are nearly identical.

The uncertainty on $P$ is 2.8 s (90\% confidence), found
from Monte Carlo simulations. 
We generated 10,000 sets of simulated LCs, each like the sets of
the actual data points. In doing that, we used the best-fit parameters
and added to each set of LCs Gaussian-distributed deviates, where
the Gaussian distribution was estimated from the residuals to
the best-fit model. Having standard deviation $\sigma =0.04$ mag, 
the Gaussian mimics the relatively large scattering of the data points from
the best-fit model.
We then fit each set of
simulated LCs with a sinusoidal function. The uncertainty on $P$
was determined by the spread of values. 
We also determined the uncertainty on the phase this way, and
found it to be 0.008 (90\% confidence).
Comparing to the X-ray ephemeris (phase at MJD 54599.0 is $\simeq$0.6714 
with a negligible uncertainty; \citealt{har+08}), 
the optical periodicity is consistent with being orbital.  
We investigated whether the period uncertainty might be caused by 
the uncertainty on the GMOS exposure recording,
because it is not clear how accurate the latter was.
We made simulations by assigning randomly produced, uniformly distributed
time offsets to the recorded image times, and found that the period 
value is not sensitive to any possible offsets.
For example, conservatively assuming 1-s uniformly distributed offsets for 
the GMOS time recording,
the resulting period difference has a range of 0.03 s, negligible
compared to the statistical period uncertainty.

The average brightness of \saxs\  in the three nights increased
from 21.123, to 21.105, to 21.023 mag, while the semiamplitude of the modulation
decreased from 0.214, to 0.202, to 0.191 mag (Table~\ref{tab:pho}). 
These variations may suggest that the two components of the emission,
the persistent and modulated, were independent of each other; as the former
was increasing, the modulation fraction 
was decreasing.  However the uncertainties on these 
parameters are relatively large, $\sim$0.04 mag (90\% confidence), 
showing that the variations of the semiamplitude are not significant.
This is because each of our observations covered only
1.5 orbital cycles, insufficient for an accurate determination.
Therefore we conclude that we have detected an approximately 20\% 
flux modulation from \saxs\ in $r'$ band.
In addition, the optical peaks correspond to when the pulsar 
is right in front of the companion (superior conjunction of the companion;
270\arcdeg\ mean orbital longitude),
confirming the previous results from \citet{hccv01} and \citet{cam+04}.

\section{Discussion}

Using the 8-m Gemini South Telescope, we have obtained, for the first time,
well-determined 
LCs from \saxs\ in its quiescent state over a time span of four days. 
From the above studies of the LCs, we find that the optical period and 
phase are consistent with the X-ray ephemeris,
indicating that the optical modulation is orbital in origin.
In studies of several tens of LMXBs at optical wavelengths
(e.g., \citealt{vm95}), in no instance has there been an accretion disk giving
rise to a sinusoidal modulation at the orbital period.
In addition, the sinusoidal maximum must correspond to 
superior conjunction of the companion star.
Because of these, we rule out the possible disk origin for
the modulation that we have suspected. 
However, the source in outburst could still be a superhumper,
which might have been hinted in the X-ray LC (Figure~\ref{fig:xlc}). 
As the outward extension of accretion disks in outburst 
has both been observed and reproduced in disk instability simulations
\citep{osa96, dhl01}, it would not be unexpected for the accretion disk
in \saxs\ to have extended to the resonance zone during the 2002 outburst, 
developing into an eccentric form due to the tidal instability \citep{wk91}.
In fact, superhumps have been seen in outbursts of both black-hole and
NS LMXB systems \citep{oc96, ele+08}.
In order to determine this possibility for the long periodicity seen 
in the 2002 outburst, time-resolved imaging observations, like ours,
of the source in outburst are needed. Since the source 
will be as bright as $\sim$17 mag in an outburst (e.g., \citealt{ghg99};
\citealt{wan+01}), a search for superhump modulation will be feasible even 
with a small telescope.

Based on the current observational studies of LMXBs 
(e.g., \citealt{vm95}), it seems
extremely unlikely that the observed optical modulation would
arise from a source other than the companion star. Thus far, pulsar wind
heating of the companion is the only model that has
been suggested \citep{bur+03, cam+04}. The long term
spin-down rate of the pulsar has been measured, indicating a rotational energy
loss rate of 9$\times 10^{33}$ ergs s$^{-1}$ \citep{har+08}. 
This energy output, presumably in the form of a pulsar wind, would illuminate
the companion star. Assuming isotropic emission and a brown dwarf companion
\citep{bc01}, the fraction of the total energy received by the companion 
is $\sim$0.005$\eta_{\ast}(R_2/0.13\ R_{\sun})^2$, where $\eta_{\ast}$ 
is the fraction of the received energy absorbed by the companion.
Following \citet{ak93}, the companion's heated face would have
temperature $\sim$$7430 \eta_{\ast}^{1/4}$ K, due to pulsar wind heating by
the putative rotation-powered pulsar.  Using such a hot face that
varies following a function of $[1+\sin i\sin (2\pi t/P)]$, 
where $i$ is the inclination angle of the binary, 
and also including a constant flux component $F_C$, 
we tested whether we could re-generate the averaged LCs of \saxs.
The distance and extinction to the source were fixed at 3.5 kpc and $A_V=0.73$,
respectively, where the extinction value is estimated from 
$A_V=N_{\rm H}/0.179\times 10^{22}$~cm$^{-2}$ by assuming
hydrogen column density to the source $N_{\rm H}=0.13\times 10^{22}$ cm$^{-2}$
(\citealt{dl90}; \citealt{hei+07}). The extinction law for 
Sloan filters given by \citet{sfd98} was used. We found that the parameter 
values of $i\simeq 63\arcdeg$ ($M_2\simeq 0.049 M_{\sun}$), 
$\eta_{\ast}\simeq 0.46$, and $F_C\simeq 19\ \mu$Jy can provide 
the observed modulation 
(the resulting $\chi^2\simeq 2100$, with no systematic uncertainties 
considered).
Although we used a very simple model, these derived 
parameter values are consistent with its known properties. 
In addition to the fact
that the companion is likely a $\sim$0.05 $M_{\sun}$ star, 
the source shows no X-ray eclipses or dips, implying $i\leq 70\arcdeg$. 
The obtained $\eta_{\ast}$ values are within the range found 
for two binary radio 
pulsars \citep{sta+01,rey+07}, in which it is known that the companion
is irradiated by the pulsar wind.
Therefore, it is plausible that the NS in \saxs\ does turn into a 
rotation-powered pulsar in quiescence, giving rise to the optical modulation.
We note that very recently, \citet{del+08} used an advanced model to 
fit their $g'i'$ light
curves, and also found that the required heating energy should 
be $\sim 10^{34}$ ergs s$^{-1}$, consistent with the derived spin-down
luminosity (which has 30\% uncertainty; \citealt{har+08}).

The origin of the persistent optical emission is not clear. \citet{hccv01}
tried explaining the emission from an X-ray irradiated disk around the pulsar,
but it may not be appropriate to use a steady thin disk model to describe
a disk in the thermally stable cold state (lower cold branch 
of the standard thermal equilibrium $S$-curve; e.g., \citealt{las01}),
since a disk temperature profile in the cold state can be drastically
different from the hot state (the steady disk case).
\citet{cam+04} used a shock front, arising from the interaction between
the companion star and pulsar wind, 
and the irradiated companion to account for the emission.
Here we argue that the accretion disk in quiescence exists, against
the suggestion that the disk would be evaporated by the pulsar 
\citep{bur+03, hei+08}, 
and this can be tested by monitoring \saxs\ at optical
wavelengths.

According to the standard disk instability model 
(DIM; e.g., \citealt{osa96,las01}), 
while the mass accretion rate to the NS in \saxs\  is very low
during quiescence,
$\dot{M}_{\rm acc}\leq 6.2\times 10^{-15}\ M_{\sun}$ yr$^{-1}$
(estimated from the observed X-ray flux), the average mass transfer 
rate from the companion to the accretion disk is as high as 
$\sim$10$^{-11} M_{\sun}$ yr$^{-1}$ 
(estimated from the X-ray fluence in each outburst; \citealt{gal08}). 
The transferred mass is stored in the disk, building up the surface 
density for triggering the next outburst. 
The average persistent $r'$ flux from \saxs\ in our observations is estimated
to be $F_C=19\ \mu$Jy, corresponding to a disk luminosity of 
$L_{r'}= 2\pi D^2 F_C/\cos i \simeq 1.5\times 10^{32}$ ergs s$^{-1}$
($i=63\arcdeg$ is assumed). 
There is plenty of gravitational energy available to power 
this emission as matter moves inwards through the outer disk.
At the time average accretion rate of 
$\dot M\sim 10^{-11}\ M_\odot \ {\rm yr^{-1}}$, matter  
falling into a radius of 4000 km releases gravitational energy at a  
rate that matches the observed luminosity. 
This radius is far larger than those that are suggested for the
inner radius $r_{\rm in}$ of the disk.
Generally, $r_{\rm in}$ would be close to the Alfv\'{e}n radius,  
$r_{\rm in}\simeq 56\ {\rm km}$ 
$(\dot{M_{\rm in}}/10^{-11} M_{\sun} {\rm yr}^{-1})^{-2/7}
(\mu/10^{26}\ {\rm G\ cm} ^{3})^{4/7}$,
where $\dot{M_{\rm in}}$ is the mass accretion rate in the inner edge
of the disk and $\mu$ is the magnetic moment, $\mu\simeq 10^{26}$ G cm$^3$ for 
\saxs\ (\citealt{har+08}). 
In the cold state, $\dot{M}_{\rm in}$ would be lower than $\dot{M}$, 
and we note that for $\dot{M}_{\rm in}= 0.1 \dot{M}$ \citep{dhl01}, 
$r_{\rm in}$ is 110 km. 
However, since a radio pulsar presumably would have no interactions 
with a surrounding disk, $r_{\rm in}$ would be larger than the light cylinder 
radius of the pulsar, which is 120 km. As can be seen, it is possible
that in quiescence, the disk in \saxs\ would be outside of the light cyliner.
In addition, the disk temperature profile in quiescence may be described by
a constant, at least right after an outburst (e.g., \citealt{osa96, dhl01}).
For \saxs, we find that an effective temperature of 4600~K for the disk
can give rise to
the persistent $r'$ flux, where the disk is assumed to be cut off
at the tidal radius 3.7$\times 10^{10}$ cm ($\simeq 0.9R_1$, where $R_1$
is the NS's Roche lobe radius).
This temperature value is consistent with those typically considered in the DIM 
(\citealt{las01}; the critical effective temperature for having
an outburst is $\sim 6000$ K). 

In order to verify our suggestion that the persistent optical emission 
arises from the disk, long-term, multi-wavelength optical monitoring 
of the source in its quiescent period is required.
From such observations, we might expect to see an increasing flux from 
the source. Moreover,
since in the DIM the temperature profile as a function of disk radius
is predicted to be changing, turning from a constant right 
after an outburst to a power-law--like function prior to the next 
outburst (e.g., \citealt{dhl01}), we would also see flux spectrum 
changes. This type of
well-behaved changes would not be expected from the pulsar wind 
shock model \citep{cam+04}, thus allowing to determine the origin
of the persistent emission. 

If the companion star is irradiated by the pulsar wind,
there is no reason to think that the disk is not. 
It has been suggested that the disk in quiescence might be 
evaporated by the pulsar (e.g., \citealt{bur+03, hei+08}),
but according to the recent calculations
by \citet{jon07}, a pulsar wind may only be effective in heating a disk.
Basically, as X-rays from a NS would ionize the surface
of a disk, the Poynting flux, which is dominant in a wind when it is
not far from the light cylinder of the pulsar, would
interact with the ionized particles, converting energy into disk heating.
Using equation~(16) in \citet{jon07}, we estimate that the baryon loss rate
of the disk at the inner radius is approximately 
3$\times 10^{21}(r_{\rm in}/120\ {\rm km})^{-3}$ cm$^{-2}$ s$^{-1}$,
only 0.05\% of the surface density ($\sim$10--100 g cm$^{-2}$) 
that is generally considered in the accretion disk 
models (e.g., \citealt{dhl01}).
This suggests that the disk in \saxs\ could exist and might be irradiated by
the pulsar wind. However, using the model provided by \citet{jon08},
the flux due to pulsar wind heating would be 2 $\mu$Jy for  
parameter $\zeta=0.3$ (0.03$\lesssim\zeta\lesssim$0.3 and a 
larger $\zeta$ value corresponds to a higher 
disk effective temperature; see details in \citealt{jon08}). 
The flux would be 10\% of the average 
$r'$ flux, which would suggest a weak pulsar-wind heating effect in \saxs.

Finally, it will be of great interest if \saxs\ can be determined to
become rotation-powered during quiescence.  
We note that the source
could be very similar to PSR~J2051$-$0827 \citep{sta+96}, a binary millisecond
pulsar system. For example, the latter has an orbital period of 2.38 hr 
and a mass function of 1.0$\times 10^{-5}$ $M_{\sun}$, and the pulsar
has a spin-down luminosity of 6$\times 10^{33}$ ergs s$^{-1}$.
However, searches for pulsed radio emission from \saxs\ have not 
been successful (e.g., \citealt{burgay+03}). Here we suggest
that the source might be identified by searching for its 
pulsed $\gamma$-ray emission. Observations of millisecond pulsars
suggest that their efficiency at $\gamma$-ray energies may be as high
as $\sim$7\% \citep{kui+00}. This implies a $\gamma$-ray flux 
of $\sim$5$\times 10^{-13}$ ergs cm$^{-2}$ s$^{-1}$ for \saxs, 
possibly detectable
by deep observations with \textit{Fermi Gamma-Ray Space Telescope}.

\acknowledgements
We thank Ian Dobbs-Dixon for useful discussion.
The Gemini queue mode observations were carried out under 
the program GS-2008A-Q-48.
The Gemini Observatory is operated by the Association of Universities 
for Research in Astronomy, Inc., under a cooperative agreement with 
the NSF on behalf of the Gemini partnership: the National Science Foundation 
(United States), the Science and Technology Facilities Council 
(United Kingdom), the National Research Council (Canada), CONICYT (Chile), 
the Australian Research Council (Australia), CNPq (Brazil), and CONICET 
(Argentina).
This research was supported by NSERC via a Discovery Grant
and by the FQRNT and CIFAR.  AC is an Alfred P.~Sloan Research Fellow.
VMK holds a Canada Research Chair and
the Lorne Trottier Chair in Astrophysics \& Cosmology, and is a
R. Howard Webster Foundation Fellow of CIFAR.

{\it Facility:} Gemini:South

\bibliographystyle{apj}

\begin{deluxetable}{l c c}
\tablecolumns{3}
\tablewidth{8cm}
\tablecaption{Photometry of J1808.4\label{tab:sax}}
\tablehead{
\colhead{MJD\tablenotemark{a}}  & \colhead{$r'$} & 
\colhead{$\Delta r'$}\tablenotemark{b}}
\startdata
0.257907 & 21.028 & 0.006 \\
0.261323 & 21.116 & 0.006 \\
0.264715 & 21.069 & 0.006 \\
0.268084 & 21.148 & 0.009 \\
0.271469 & 21.267 & 0.010 \\
\enddata
\tablenotetext{a}{Days since MJD 54597.0.}
\tablenotetext{b}{1$\sigma$ uncertainty resulting from PSF fitting.}
\tablecomments{Table~1 is published in its entirety in the electronic
edition of the Astrophysical Journal. A portion is shown here for guidance
regarding its form and content.}
\end{deluxetable}

\begin{deluxetable}{l  c c c }
\tablecolumns{4}
\tablewidth{8cm}
\tablecaption{Summary of brightnesses of nearby stars and \saxs\ in our observations
\label{tab:pho} }
\tablehead{
\colhead{Source} & \colhead{Obs 1} & \colhead{Obs 2} & \colhead{Obs 3} \\
\colhead{} & \colhead{(MJD 54597)}  & \colhead{(MJD 54598)} & \colhead{(MJD 54601)}
}
\startdata
Star $a$ & $21.545\pm0.029$ & $21.444\pm0.020$ & $21.486\pm0.021$ \\
Star $b$\tablenotemark{a} & \multicolumn{3}{c}{21.133$\pm$0.013} \\
Star $C$\tablenotemark{a} & \multicolumn{3}{c}{21.178$\pm0.013$} \\
\cutinhead{Sinusoidal fitting} 
\sidehead{J1808.4} 
Average magnitude\tablenotemark{b}   & 21.12 & 21.11 & 21.02 \\
Semiamplitude\tablenotemark{b}   & 0.21 & 0.20 & 0.19 \\
\enddata
\tablenotetext{a}{Average magnitude is derived from all three observations.}
\tablenotetext{b}{Uncertainties (90\% confidence) are $\sim$0.04 mag.}
\tablecomments{Uncertainties of 0.025 mag and probable 0.03 mag 
from the aperture correction and zero point calibration, respectively, 
are not included.}
\end{deluxetable}

\end{document}